\documentclass[Crown,sageh,times]{sagej}
\usepackage{moreverb,url}

\usepackage[colorlinks,bookmarksopen,bookmarksnumbered,citecolor=red,urlcolor=red]{hyperref}

\newcommand\BibTeX{{\rmfamily B\kern-.05em \textsc{i\kern-.025em b}\kern-.08em
T\kern-.1667em\lower.7ex\hbox{E}\kern-.125emX}}

\def\volumeyear{2016}
\RequirePackage{amsthm,amsmath,amsfonts,amssymb}
\RequirePackage{natbib}[authoryear]
\RequirePackage{hyperref}[colorlinks,citecolor=blue,urlcolor=blue]

\RequirePackage{graphicx}

\theoremstyle{plain}

\theoremstyle{remark}

\usepackage{amsmath}

\usepackage{amssymb}
\usepackage{comment}

\def\log{\hbox{log}}

\def\bse{\begin{eqnarray*}}
\def\ese{\end{eqnarray*}}
\def\be{\begin{eqnarray*}}
\def\ee{\end{eqnarray*}}
\def\bq{\begin{equation}}
\def\eq{\end{equation}}
\def\bse{\begin{eqnarray*}}
\def\ese{\end{eqnarray*}}

\def\trans{^{\rm T}}

\def\bZ{{\mathbf Z}}

\def\bW{{\mathbf W}}

\def\bX{{\mathbf X}}

\def\b0{{\mathbf 0}}

\newcommand{\balpha}{\mbox{\boldmath $\alpha$}}

\newcommand{\btheta}{\mbox{\boldmath $\theta$}}

\begin{document}

\runninghead{Luan et al.}

\title{Scalable approaches to adjusting for biases due to measurement error in multi-level generalized functional linear regression models with heteroscedastic measurement errors}

\author{Yuanyuan Luan\affilnum{1*}, Roger S. Zoh\affilnum{1}, Erjia Cui\affilnum{2}, Xue Lan\affilnum{3}, Sneha Jadhav\affilnum{4}, and Carmen D. Tekwe \affilnum{1*} }

\affiliation{\affilnum{1}Department of Epidemiology and Biostatistics, Indiana University, Bloomington, IN 47405, United States\\
\affilnum{2}Department of Biostatistics, Johns Hopkins University, Baltimore, MD 21205, United States \\
\affilnum{3}  Department of Statistics, Oregon State University, Corvallis, Oregon, United States\\
\affilnum{4} Department of Mathematics \& Statistics, Wake Forest University, NC, 27109, United States}

\email{ctekwe@iu.edu}

\begin{abstract}
Wearable devices permit the continuous monitoring of biological processes, such as blood glucose metabolism, and behavior, such as sleep quality and physical activity. The continuous monitoring often occurs in epochs of 60 seconds over multiple days, resulting in high dimensional longitudinal curves that are best described and analyzed as multi-level functional data. From this perspective, the functional data are smooth, latent functions obtained at discrete time intervals and prone to homoscedastic white noise. However, the assumption of homoscedastic errors might not be appropriate in this setting because the devices collect the data serially. While extensive methods exist for measurement error correction associated with scalar covariates prone to errors, less work has been done on adjusting for biases due to measurement error in high dimensional longitudinal curves prone to measurement errors. We present two methods for adjusting for biases due to measurement error in longitudinal functional curves prone to complex measurement error structures in multi-level generalized functional linear regression models. These two-stage-based methods employ functional mixed effects models. We assume that the distribution of the scalar responses and the functional observed measures prone to heteroscedastic errors both belong in the exponential family and relax the assumption on the distribution of the measurement error. Through simulations, we established some finite sample properties of these methods. The methods were applied to assess the relationship between device-based measures of step counts and type 2 diabetes in community-dwelling adults living in the United States who participated in the National Health and Nutrition Examination Survey.
\end{abstract}


\maketitle

\section{INTRODUCTION}\label{section:intro}
Wearable monitoring devices permit the continuous monitoring of biological processes, such as blood glucose metabolism \citep{tsai2019diabetes,gaynanova2022modeling}, and behavior, such as sleep quality \citep{kuo2016development,tuominen2019sleep} and physical activity \citep{troiano2008physical}. The continuous monitoring often occurs in epochs of 60 seconds over multiple days resulting in high dimensional multi-level longitudinal curves that are best described and analyzed as multi-level functional data. 

While researchers have previously addressed measurement error in scalar covariates prone to error \citep{carroll:2006, fuller:2009}, less work has been done on correcting measurement error in multi-level high dimensional curves prone to heteroscedastic measurement errors \citep {cardot2007smoothing,crambes2009smoothing,goldsmith2011penalized}. When correcting for measurement error, most approaches assumed that the serially observed functional data represent a curve of a latent process contaminated by noise with an independent error structure \citep{silverman2005,yao2005functional,rice2001nonparametric}. For example, Yao and colleagues\citep{yao2005functional} proposed an approach, named principal components analysis through conditional expectation (PACE), to incorporate uncorrelated and additive measurement error with a common variance, which extended the classical functional principal components (FPC) by framing the FPC scores as conditional expectation. They assumed that the true measurements were independent realizations of a smooth function with an unknown mean function, $\mu(t)$, and a covariance function. They also assumed a discrete measurement error, denoted by $U_{ij}$ with mean 0 and a common variance. The model they proposed is $W_{ij} = X_i(T_{ij}) + U_{ij}$, where $W_{ij}$ is the 
$jth$ observation of the random function $X_i$ (at time $T_{ij}$) that is contaminated with measurement error $U_{ij}$. However, because observation periods for functional data often extend over multiple days, it is necessary to develop methods for correcting measurement errors in serially observed data prone to more complex heteroscedastic structures. Other researchers have proposed approaches to overcome these limitations. Crainiceanu et al. \citep{crainiceanu2009generalized} demonstrated that the classical measurement error model can be generalized to a multilevel functional mixed effect model, where the repeated measured observation $W_{ij}(t)$ is considered to be a proxy observation of the true measurement $X_i(t)$ that comprise of visit-specific deviation, subject-wise deviation, measurement error, and random noise. Additionally, prior approaches to correcting measurement error in serially observed functional data prone to complex errors involved the assumption that the observed curves were from the Normal distribution \citep{tekwe2022estimation,jadhav2022function}. However, sometimes the curves might be best described as belonging to more flexible distributions such as those in the exponential family \citep{crainiceanu2009generalized}. To address these limitations, we propose two mixed effects-based methods for bias correction due to measurement error in multi-level functional data from the exponential family of distributions and prone to complex heteroscedastic measurement error. 

Our current work was motivated by the need to assess the role of device-based measures of step counts in type 2 diabetes (T2D). The incidence and prevalence of T2D have increased over the past 20 years. Over one-third of adults in the United States are obese, and $\ge$ 11\% of people over 20 years old have T2D \citep{eckel2011obesity}. Some researchers project that the overall prevalence of T2D in the United States will be 21\% by 2050 \citep{boyle2010projection}. Obesity has been associated with a chronic imbalance between energy intake and energy expenditure \citep{spiegelman2001obesity}. Thus, to combat the growing epidemic of obesity and its associated health outcomes such as T2D, clinicians and researchers have proposed treatments to reduce dietary intake and increase daily physical activity \citep{balk2015combined}. Therefore, it is critical to assess these behaviors accurately and evaluate how they influence obesity and T2D. Self-reported measures of physical activity are prone to recall bias \citep{sallis2000assessment,armstrong2006development,tooze2006new}. Consequently, researchers increasingly use wearable devices to monitor physical activity \citep{kozey2010accelerometer}. Wearable devices for physical activity are based on motion sensor technology and data processing to provide measures of physical activity frequency, duration, and intensity \citep{kozey2010accelerometer,troiano2008physical}. While wearable device measures of physical activity are not prone to recall and researcher bias, their accuracy may be questionable \citep{0967-3334-33-11-1769, crouter2006estimating, jacobi2007physical,warolin2012,rothney2008validity,kozey2010accelerometer}. The sources of error in wearable device measures of physical activity are both random and systematic. To process accelerometer data from wearable devices, the devices must first be calibrated to record activity counts with another physiological variable, such as the metabolic equivalent task (MET) \citep{kozey2010accelerometer,crouter2006estimating,swartz2000estimation}. Next, the device approximates activity counts from its relationship with the physiological variable with a regression equation \citep{kozey2010accelerometer}. However, there are currently no standard equations, with over 30 equations available for this estimation \citep{plasqui2007physical,kozey2010accelerometer,crouter2006estimating,rothney2008validity}. The accuracy of the data generated by the devices thus depends on the estimation equations. The accuracy also depends on the activity type, sex, and body composition of device users \citep{freedson1998calibration, valenti2014validating}. In short, physical activity is a latent variable not observed directly but represented by a proxy, such as wearable device-based measures. Such observed measurement tends to introduce measurement error \citep{carroll:2006}.

In this work, we first develop mixed effects-based methods to adjust for biases due to the presence of measurement error in multi-level generalized functional regression models. To do this, we assume the distributions of the functional covariates prone to measurement errors belong in the exponential family. This assumption allows for a more general specification of the distributions of error-prone functional covariates compared to current approaches that often entail normality assumptions for these observed measures. By this, we allow the true measurement and observed measurement prone to measurement error to have a non-linear association. Second, we treat the random errors in the observed measures as complex heteroscedastic errors from the Gaussian distribution with covariance error functions. Third, our methods can be used to evaluate relationships between multi-level functional covariates with complex measurement error structures and scalar outcomes with distributions in the exponential family. Fourth, we treat the functional covariate as an observed measure for the true functional unobserved latent covariate. Additionally, our proposed methods employ a point-wise method to fit the multi-level functional mixed effects-based approach previously described by Cui et al \citep{cui2021fast}, avoiding the need for computing complex and intractable integrals which would be required in traditional approaches to reducing biases due measurement error in multi-level functional data. We perform extensive simulations to compare our proposed methods with current approaches to measurement error in functional data and other methods that do not perform any adjustments for measurement errors. We report on our application of the proposed methods to assess the relationship between device-based measures of physical activity and T2D in data from the National Health and Nutrition Examination Survey (NHANES).

\section{GENERALIZED FUNCTIONAL REGRESSION WITH MULTI-LEVEL HETEROSCEDASTIC MEASUREMENT ERRORS} \label{section:models}

\subsection{The model} 
Let $\mathcal{D}_i = \{Y_i,X_i(t),\bZ_i\}$ be a triplet for individual $i (i=1,\dots, n)$ consisting of, respectively, a scalar outcome $Y_i$, a random functional covariate $X_i(t)$ assumed to be square integrable on the unit interval, and $\bZ_i \in \mathbb{R}^{p}$ is a vector of length $p$ representing the error-free covariates. The model for the $i${th} subject is:
\begin{eqnarray}
g[E\{Y_{i} \mid X_{i}(t), \bZ_{i}\}] &=& \int_{0}^{1} \beta(t)X_{i}(t)  dt +  \bZ_i\trans\balpha,\label{eq1}\\
h[E\{W_{ij}(t) \mid X_i(t)\}] &=& X_i(t),  \label{eq2} \\
X_i(t) &=& \mu_x(t) + \varepsilon_{xi}(t);\label{eqX} 
\end{eqnarray}
 where $g(\cdot)$ and $h(\cdot)$ are both known monotone, twice continuously differentiable functions, and $\beta(t)$ is an unknown functional coefficient associated with $X_i(t)$. The response variable $Y_i$ may be either a continuous or discrete outcome with a distribution belonging to the exponential family, and $\balpha$ is a $p\times1$ vector of coefficients associated with the $p$ error-free covariates, $\bZ_i$. Furthermore, we assume that $X_{i}(t)$ is not directly observable, instead a set of $J$ surrogate functional variables $\left\{W_{ij}(t)\right\}_{j=1}^{J}$ are observed for the true functional covariate $X_{i}(t)$ as in (\ref{eq2}). The measurement error model in (\ref{eq2}) is determined by the exponential family distribution assumed for $W_{ij}(t)$. We note that the exponential family specification of the measurement error component of the model is a significant departure from current approaches that are based on Gaussian processes and classical measurement error models. Lastly, the random term, $\varepsilon_{xi}(t)$ in Equation \ref{eqX} is the subject-specific deviation of each $X_i(t)$ from overall the mean process, $\mu_x(t)$, and we assume $\varepsilon_{xi}(t) \sim GP\{\boldsymbol{0}, \Sigma_{xx}(t,t')\}$.
 
 In our motivating example, $Y_i$ represents a health outcome such as T2D status, $X_i(t)$ is the true overall daily physical activity for individual $i$, which is approximated by device-based measures such as step counts measured over $J$ distinct days, denoted by $\bW_i = \{W_{ij}(t)\}_{j=1}^{J}$.

%\in \mathbb{R}^{T_i \times J},{t \in T_{i}}$ which is measured at multiple time points, where $T_{i}$ is the number of distinct time points at which $W_{ij}(t)$ are measured . 

 %$W_{ij}(t)$ is the $t${th} $t=(1,\ldots,T)$ observation collected at the $j${th} $j=(1,\ldots,J)$ day for subject $i$.  We assume that $\beta(t)$ is a smooth function that can be approximated by spline functions. In our motivating example, $Y_i$ represents a health outcome such as T2D status, $X_i(t)$ is true overall physical activity which is approximated by device-based measures of physical activity such as step counts, $\bW_i = \{W_{ij}(t)\}_{j=1,\cdots,J} \in \mathbb{R}^{T_i \times J},{t \in T_{i}}$ which is measured at multiple time points, where $T_{i}$ is the number of distinct time points at which $W_{ij}(t)$ are measured. The random term, $\varepsilon_{xi}(t)$, in Equation \ref{eqX} is the subject-specific deviation of each $X_i(t)$ from overall the mean process, $\mu_x(t)$, and we assume $\varepsilon_{xi}(t) \sim GP\{\boldsymbol{0}, \Sigma_{{\varepsilon}{\varepsilon}}(t, t')\}$. The measurement error is determined by the exponential family distribution assumed for $W_{ij}(t)$. We note that the exponential family specification of the measurement error component of the model is a significant departure from current approaches that are based on Gaussian processes and classical measurement error models. 
 
\section{MODEL ASSUMPTIONS AND ESTIMATION} \label{section:estimate}

Estimation and inference about the functional parameter $\beta(t)$ is our primary goal, but $X_i(t)$ is latent and not directly observed. Estimating $\beta(t)$ with the observed measure of $X_i(t)$, $W_{ij}(t)$, leads to biased estimates of $\beta(t)$ \citep{carroll:2006, tekwe2022estimation, tekwe2017functional, tekwe2016generalized,cardot2007smoothing}. The biased estimator of $\beta(t)$ may attenuate or over-estimate the true effects of $X_i(t)$ depending on the regression model. However, the presence of some additional information in the data, such as validation data, repeated measurements on $W_{ij}(t)$, or instrumental variables for $X_i(t)$, may be used to adjust for the measurement errors \citep{carroll:2006}. We use replicates for model identification and propose two approaches based on extensions of functional mixed effects models \citep{guo2002functional} for adjusting for biases due to the presence of measurement errors in the multi-level functional data.

Prior to the estimation of the model parameters, we make the following assumptions:
 
\begin{itemize}
\item[[A1]] $Y_i\mid X_{i}(t), \bZ_{i} \sim EF(\cdot)$, where EF refers to an exponential family distribution.
\item[[A2]] The functions, $g(\cdot)$ and $h(\cdot)$, are known monotone, twice continuously differentiable functions.
%\item[[A3]] $\mathrm{Cov}(\bX_i,\bU_i)= \b0$.
\item[[A3]] $\mathrm{h[{E}}\{W_{ij}(t) \mid X_i(t)\}] = X_i(t)$
%and $\mathrm{Cov}(\bW_i \mid \bX_i) = \Sigma_{uu}$.
\item[[A4]] $\mathrm{Cov}\{W_{ij}(t),X_i(t)\} \neq 0$.
%\item[[A5]] $\mathrm{Cov}\{X_{i}(t),X_{i}(t')\} \neq 0$ and $\mathrm{Cov}\{W_{ij}(t),W_{ij}(t')\} \neq 0$ for $t \ne t'$.
\item[[A5]] $W_{ij}(t)|X_i(t) \sim EF(\cdot)$.
\item[[A6]] $f\{Y_i|X_i(t),W_{ij}(t)\}= f\{Y_i|X_i(t)\}$.
%\item[[A7]] $W_{ij}(t) = E\{W_{ij}(t)|X_i(t)\} + U_{ij}(t)$
%\item[[A8]] $U_{ij}(t) \sim GP(0, \Sigma_{uu}(t,t'))$,  $\mathrm{Cov}\{U_{ij}(t),U_{ij'}(t)\} \neq 0$ for $j \ne j'$, and $\mathrm{Cov}\{U_{ij}(t),U_{ij}(t')\} \neq 0$ for $t \ne t'$.
\item[[A7]] $X_{i}(t) \sim GP\{\mu_x(t), \Sigma_{xx}(t,t')$\}. %We do not need to specify a particular distribution of $X(t)$ Because the random effects in a generalized mixed-effect model do not have a specific distribution rather than just a common multivariate distribution F() \citep{analysis of longitudinal data P129}
\end{itemize}

A1 indicates that the scalar response may be discrete or continuous with a distribution belonging to the EF. A2 includes the usual assumptions for link functions in generalized linear regression models. A3 indicates that nonlinear functions of $\mathrm{h[{E}}\{W_{ij}(t) \mid X_i(t)\}]$ are unbiased measures of $X_i(t)$. It specifies a non-linear association between the true and observed measures. The corresponding measurement error model is $W_{ij}(t) = E\{W_{ij}(t)|X_i(t)\} + U_{ij}(t)$, which is generalized from the classic additive measurement error model $W = X + U$, where the true measures $X$ and observed measures $W$ have a linear relationship. In other words, the classic additive measurement error is a special case of the proposed measurement error model, when $h(\cdot)$ is the identity function. A4 states that the observed measures and the true unobserved covariate are correlated, a classical assumption in measurement error models. A5 indicates that observed measures belong in the EF. While the non-differential measurement error assumption specifying that the observed measure, $W_{ij}(t)$, does not provide any additional information about the response, $Y_i$, beyond the information given by $X_i(t)$ is given in A6. %A8 indicates that subject- and repeated session-specific deviation of the observed measures from the true covariates have a Gaussian process. Additionally, the covariance function for $U_{ij}(t)$ is allowed to be correlated across the $J$ days of observation and also across the wear times, $t$. 
A7 indicates that the true measures follow a Gaussian process with a mean function, $\mu_x(t)$, and a covariance function, $\Sigma_{xx}(t,t')$. This is a typical assumption for the measurement error model. In addition, the proposed model allows for correlated measurement errors for the observed measures and the true covariate for each subject $i$. The proposed model also relaxed the assumption on the distribution of measurement error.

%\subsection{Likelihood}
Given the observed data, $\mathcal{D}_i = \{Y_i,\bW_i,\bZ_i\}$, and identifying data for model identification, we propose the following likelihood, $\ell_{i}$, for each subject $i$,
\begin{eqnarray}
\ell_{i}\{Y_i, \bX_{i}, \beta(t), \balpha, \btheta_w\} &=& f_{y}(Y_i|\beta(t), \balpha, \bX_{i}, \bZ_i)\prod^{J}_{j=1}f_{W}(\bW_{ij}|\bX_{i},\btheta_w), \label{eq:eqliki}
\end{eqnarray}
where $f_{y}(\cdot)$ and $f_{w}(\cdot)$ are densities from the exponential family which depend on the choice of the link function and the random component. We note that the likelihood specified in Equation~\ref{eq:eqliki} is flexible as it allows for non-normal data. For example, with an identity link function that has Gaussian random components for $h(\cdot)$ and $g(\cdot)$, our model becomes the usual Gaussian measurement error model. The joint likelihood then is 
\begin{eqnarray}
\ell &=& \prod^{n}_{i=1} \ell_{i}\{Y_i,\bX_{i}, \beta(t), \balpha, \btheta_w\}. \label{eq:eqlik}
\end{eqnarray}

In likelihoods \ref{eq:eqliki} and \ref{eq:eqlik}, the functional covariate $\bX_i = \{X_{i}(t)\}^{i \in n}$ is a random function because it is unknown and unobserved, which is characteristic of the scalar-on-function regression with measurement error \citep{tekwesim}.

We discuss the functional mixed effects-based approaches to bias correction of our proposed models in the next section. 

%For inference, we propose the use of nonparametric point-wise confidence intervals. 

\section{Mixed Effects Model-Based Approach to Measurement Error Correction}

Random or classical measurement error in the exposure variable occurs when the observed measure for the true covariate is distributed around the true exposure \citep{nab2021sensitivity}. Validation data or instrumental variables are useful for correcting classical measurement error in estimating model parameters. These additional identifying data are used in measurement error correction methods such as regression calibration \citep{carroll:2006,spiegelman1997fully,wang1997regression}. However, such data are rarely available \citep{thurigen2000measurement} in most epidemiological studies. Additionally, implementing the traditional regression calibration method for multi-level high-dimensional data such as functional data can be computationally intensive and require the evaluation of intractable high-dimensional integrals. To avoid these potential computational complexities, we propose a mixed effect model-based (MEM) approach to measurement error correction, which is a two-stage approach to estimating the parameters in Equation \ref{eq1}. Under the first stage, we obtain measurement error adjusted estimates for the $X_i(t)$, denoted as $\hat{X_i(t)}$, which are then plugged into the outcome models to assess the associations of the error-prone and error-free covariates on the outcomes (second stage). Since the second stage mainly involves regular functional regression analysis, which is not the focus of this manuscript, we will focus on describing the first stage of the approximation process of $X_i(t)$ in this section. 

Under assumption A3 in Section \ref{section:estimate}, we assume that $\mathrm{h[{E}}\{W_{ij}(t) \mid X_i(t)\}] = X_i(t)$. Thus, these observed measures are assumed to be unbiased observed measures for the true unobservable latent covariate. Using the repeated observations on the unbiased measures to obtain $\hat{\mathrm{h[{E}}\{W_{ij}(t) \mid X_i(t)}\}]$ allows us to obtain expressions for $\hat{X_i(t)}$. This will be accomplished by developing measurement error adjustment approaches to multi-level generalized functional linear regression by reframing the measurement error model to generalized functional linear mixed effects \citep{schipper2008generalized,scheipl2016generalized,scheipl2015functional} framework. To do this, we first re-write the measurement error model (Equation \ref{eq2}) in a generalized functional linear mixed-effect model format as follows: 
\begin{eqnarray}
h[E\{W_{ij}(t)|\gamma_{i}(t)\}] &=& \mu_x(t) + \gamma_{i}(t) \label{UPMEM},
\end{eqnarray}
where $\mu_x(t)$ is the functional fixed intercept and $\gamma_{i}(t)$ is the functional random intercept. We assume that $\gamma_{i}(t) \sim GP\{\boldsymbol{0}, \boldsymbol{G}(t, t')\}$. The $\mu_x(t)$ is the overall population mean process of $h\{W_{ij}(t)\}$, which is $h[E\{W_{ij}(t)\}]$. $\gamma_{i}(t)$ is the subject-specific deviation of $h\{W_{ij}(t)\}$ from the overall population mean process, $h[E\{W_{ij}(t)\}]$. By re-writing measurement error model (Equation \ref{eq2}) in the generalized functional linear mixed-effect model  format (Equation \ref{UPMEM}) while combining it with the probability distribution of the true measurement $X_i(t)$ (Equation \ref{eqX}), we obtain:
\begin{eqnarray}
h[E\{W_{ij}(t)|X_{i}(t)\}] =X_{i}(t) &=&h[E\{W_{ij}(t)|\gamma_{i}(t)\}] =  \mu_x(t) + \gamma_i(t) \label{eqWX}.
\end{eqnarray}

Therefore, it is reasonable to approximate the true measurement, $X_i(t)$, with $h[E\{W_{ij}(t)|\gamma_{i}\}]$ using a generalized functional linear mixed-effect model. Throughout this manuscript, we will use $h[E\{W_{ij}(t)|\gamma_{i}(t)\}]$ and $h[E\{W_{ij}(t)|X_i(t)\}]$ interchangeably.

\subsection{Estimation and Inference}
Applications of the MEM approach to measurement error adjustment involve an analysis of multi-level functional data and computations of potentially intractable and complex integrals with increased computational burden. We propose to avoid these potential computational complexities by first performing point-wise generalized linear mixed effects analysis of $W_{ij}(t)$ by $t$, the time of observation as described by Cui et al. \citep{cui2021fast}. We label this approach univariate point wise MEM (UP\_MEM). This first step yields approximated values for $X_{i,t}$, $h[\hat{E}\{W_{ij, t}|X_{i, t}\}]$, obtained from the pointwise univariate approach to fitting the model for the response variable in Equation \ref{UPMEM} at each unique time point or wear time across the repeated days of observation. %In the second step, we smooth the approximated values of the true measurement $X_i(t)$, $h[E\{W_{ij}(t)\}|X_i(t)]$ across $t$ with any smoother, such as B-splines or penalized splines. 
The use of generalized linear mixed effects models to obtain the approximated values for $X_i(t)$ is possible due to the repeated measures on $W_{ij}(t)$ across the $J$ days. In the second step, we substitute these values for $X_i(t)$ in the outcome regression equation in Equation \ref{eq1}. The last step involves non-parametric bootstraps to estimate confidence bands. 

 A potential drawback of the UP\_MEM method is that it may not fully account for the serial correlation across days or observation periods adequately due to its univariate approach to analyzing the data by single time points at a time. To address this potential limitation, we propose the multivariate point-wise MEM (MP\_MEM) approach. The MP\_MEM method also involves fitting point-wise generalized linear mixed effects models to the data with Equation \ref{UPMEM}, but by analyzing observations from multiple time points concurrently and proceeding across all the $T_i$ time points in a moving average manner. The term "multivariate" here refers to the use of multiple time points or wear times of the device from our motivating example concurrently in fitting the point-wise generalized linear mixed effects model. In the simulation study, we fitted the generalized-mixed effects model at every three adjacent time points because we believe adjacent measures of device-based physical activity are more correlated with each other. However, more time points can be used simultaneously in the estimation depending on the application.

We now provide detailed descriptions of using the MEM approach\citep{guo2002functional,antoniadis2007estimation,morris2006wavelet,
scheipl2015functional} for adjusting for measurement error biases in the generalized functional linear regression models using point-wise generalized linear mixed effects analysis. The first approach, UP\_MEM\citep{cui2021fast}, was proposed by Cui et al. (2021) \citep{cui2021fast} which we adapt for addressing measurement error in multi-level functional data. The second approach, namely, the MP\_MEM method, was proposed by us. The common principle shared by the two approaches is that under the assumption that $\mathrm{h[{E}}\{W_{ij}(t) \mid X_i(t)\}] = X_i(t)$, the fitted values from regressing $h[E\{W_{ij}(t)\mid r_i(t)\}]$ in the generalized functional mixed-effect model with a fixed and random intercept (see Equation \ref{UPMEM} and Equation \ref{eqWX}) can be used as approximated values of $X_i(t)$ when assessing its association with the outcome of interest such as in Equation \ref{eq1}. For our first approach, we adapt the UP\_MEM modeling developed by Cui and colleagues \citep{cui2021fast} to obtain the approximation of the true measures for the outcome regression model at each time point (location). For our second approach, we proposed an MP\_MEM modeling to obtain the approximation of the true measures for the outcome regression model across multiple time points simultaneously, providing the ability to address potential correlations between discrete observations at different time points (location). 
The MP\_MEM method is implemented in this manuscript in the following five steps:

\begin{enumerate}
\item Given the assumed measurement error model and the conditional distribution of $W_{ij}(t)$, fit the generalized mixed-effects model at every $D$ adjacent time points on the functional continuum 
\begin{eqnarray*}
% h[E\{W_{ij, t}|r_{i, t}\}] &=& h[E\{W_{ij, t}|X_{i, t}\}]= \mu_{x_t} + r_{i,t} + , \label{UP\_MEM}
h\{E(W_{ij}|r_i)\} &=& h\{E(W_{ij}|X_i)\}= \mu_{X} + r_{0i}+ r_{1i}I(t_1) + r_{2i}I(t_2) \dots + r_{Di}I(t_D), 
\end{eqnarray*} 
where $\mu_{X}$ is the fixed intercept or overall mean for $X_i(t)$ across the $D$ adjacent time points and $r_{0i}$ is the random intercept of the subject in the model, which represents the subject-specific deviation of $X_i$ from the overall mean $\mu_{X}$. $\{r_{1d}\}^{D}_{d=1}$ are the random effects of time, which represents the time-wise deviation of $X_i$ from the overall mean $\mu_{X}$. $t_{ij}$ is the categorical time variable for the jth time point in the short period. We set $D$ to be three in the simulation study and data application so that two adjacent data points will be utilized for obtaining estimated values for $X_i(t)$ at the given time of interest. This process is repeated to obtain approximate values for $X_i(t) $ over all the available $T$ time points, except for data at the beginning and end of the functional continuum. More time points can be used in practice if the correlations are assumed to exist beyond the immediate adjacent observed measures in the data. We obtain $h[\hat{E}\{W_{ij}(t_d)|X_{i}(t_d)\}]$, the estimated value of $h[E\{W_{ij}(t_d)|X_{i}(t_d)\}$ in this step for each $t_d \in T$, where $t_d$ is the time point of interest.   

\item Next, put $h[\hat{E}\{W_{ij}(t_d)|X_{i}(t_d)\}]$ from all the time points together in the original order of $T$, and consider it to be $h[\hat{E}\{W_{ij}(t)|X_{i}(t)\}]$, i.e., $\hat{X}_i(t)$.

\item Replace $X_{i}(t)$ with $\hat{X}_i(t) = h[\hat{E}\{W_{ij}(t)|X_{i}(t)\}]$ in the response regression model Equation \ref{eq1}. Most current approaches to correcting measurement error in functional data analysis with functional covariates prone to measurement error require some dimension reduction before correcting measurement error \citep{yao2005functional,crainiceanu2009,tekwesim,tekwe2022estimation}. An advantage of our proposed methods is that they do not require dimension reduction of $W_{ij}(t)$ to obtain predicted values of $X_{i}(t)$ for fitting the regression model in Equation \ref{eq1}. 

\item While dimension reduction approaches are not required to obtain the approximated measure of $X_i(t)$, they are, however, required in the regression step (stage 2). This is accomplished by reducing the dimensions of the functional terms in Equation \ref{eq1} as follows: 
$\beta(t)\approx \sum_{k=1}^{K_{n}}\omega_{k}b_{k}(t)$ and 
$\hat{X}_{ik} \approx
\int_{0}^{1}\hat{X}_{i}(t)b_{k}(t)dt$, 
where $\left\{\omega_{k}\right\}_{k=1}^{K_{n}}$ are unknown spline coefficients and $\left\{
b_{k}(t)\right\} _{k=1}^{K_{n}}$ are a set of spline basis functions on $[0,1]$. Equation \ref{eq1} becomes 
\begin{eqnarray}
g[E\{Y_{i} \mid \bX_{i}, \bZ_{i}\}] &\approx& \sum_{k=1}^{K_{n}}\omega_{k}\hat{X}_{ik} + \bZ_{i}\trans\balpha. \label{reduced_model}
\end{eqnarray}

\item Maximize the likelihood function in the reduced form model in Equation \ref{reduced_model} to obtain the estimated model parameters
$E\{\hat{\beta}(t)\} \approx E\{\sum_{k=1}^{K_n} \hat{\omega}_{k}b_k(t)\}$ and $E(\hat{\balpha}) = \balpha$. 
\end{enumerate}

The procedure of the UP\_MEM approach has the same steps as the MP\_MEM for Steps 2-5. However, for the UP\_MEM approach, we fit the generalized-mixed effects model at each unique time point. That is $h[E\{W_{ij}(t_d)|r_{i}(t_d)\}] = h[E\{W_{ij}(t_d)|X_{i}(t_d)\}]= \mu_{X_{t_d}} + r_{it_d}$, where $\mu_{X_{t_d}}$ is the fixed intercept or overall mean for $X_i(t_d)$ at time of interest $t_d$ and $r_{it_d}$ is the random intercept of the subject in the model, which represents the subject-specific deviation of $X_{i}(t_d)$ from the overall mean $\mu_{X}$ at time $t_d$. 

%A potential drawback of the UP\_MEM method is that it may not fully account for the serial correlation across days or observation periods adequately due to its univariate approach to analyzing the data by single time points at a time. To address this potential limitation, we propose the MP\_MEM approach. The MP\_MEM method also involves fitting point-wise generalized linear mixed effects models to the data with Equation \ref{eqUP\_MEM}, but by analyzing observations from multiple time points concurrently and proceeding across all the $T_i$ time points in a moving average manner. The term "multivariate" here refers to the use of multiple time points or wear times of the device from our motivating example concurrently in fitting the point-wise generalized linear mixed effects model. The procedure of the MP\_MEM approach has the same steps as the UP\_MEM for Steps 2-5. However, for the MP\_MEM approach, we fit the generalized-mixed effects model at a few time points simultaneously. In the simulation study, we fitted the generalized-mixed effects model at every 3 time points. More time points can used if believing there are long distance correlations. 

Our methods for adjusting for bias due to measurement error allow arbitrary error structures for $\Sigma_{uu}(t,t')$, such as unstructured (UN), compound symmetry (CS), or auto-regressive 1 \{AR(1)\}. 

For inference, we used $95\%$ nonparametric point-wise bootstrap confidence intervals.

\section{SIMULATIONS}
We conducted simulations to investigate the finite sample properties of our methods. 

We generated 1,000 data sets with sample sizes
$n \in (100, 200, 500, 1,000, 2,000, 5,000)$ independently from the model in Equation \ref{eq1} with $g(\cdot)$ being the logit function. We simulated the outcomes as $Y_i \sim Binomial(p_i)$ with $p_i = expit(\eta_i)$ and $\eta_i = \int_{0}^{1}\beta(t)X_i(t)dt + \boldsymbol{Z_i \alpha}$, where $\beta(t) = sin(2\pi t)$ and $\boldsymbol{\alpha} = (1, 2)$. We simulated the true functional covariate with $X_{i}(t)=1/[1+\exp\{8(t-0.5)\}] + \epsilon_{x_i}(t))$, and generated $\epsilon_{x_i}(t)$'s independently from a Gaussian process (GP) with mean 0 and an AR(1) structure for the covariance function. That is, $\epsilon_{x_i}(t) \sim GP\{\boldsymbol{0}, \Sigma_{xx}(t,t')\}$. $\Sigma_{xx}(t,t')$ is an autoregressive of order 1 \{AR(1)\} covariance function. $\Sigma_{xx}(t,t')=\sigma^2_{X} \rho_{X}^{|t-t'|}$, where $\sigma_{X}$ is the constant standard deviation of $X(t)$ along the functional span and $\rho_X$ is the correlation between observations of lag 1. We included a continuous error-free covariate, $Z_{1i} $, and a binary error-free covariate, $Z_{2i}$. We simulated them independently with $Z_{1i} \sim N(2, 1)$ and $Z_{2i} \sim Binomial(1, 0.6)$, respectively. We generated the observed measure, $W_{ij}(t)$, from a Poisson process (PP) with mean function $\lambda_i(t)$. That is, $W_{ij}(t) |X_i(t) \sim PP\{\lambda_i(t)\}$, where $\lambda_i(t) = \exp\{X_i(t)\}$ and $j = 1, \dots, J=5$. %The measurement error component of our model in Equation \ref{eq2} was $\log[E\{W_{ij}(t)\}]= X_i(t) + U_{ij}(t)$ for the $i${th} subject at the $j${th} replication, and we assumed $U_{ij}(t) \sim GP\{\boldsymbol{0}, \Sigma_{uu}(t,t')\}$ for $t \ne t'$. 

We compared six estimators for $\beta(t)$ in Equation \ref{eq1} in analyzing the simulated data: UP\_MEM, MP\_MEM, PACE, Average, Naive, and Oracle. The Average estimator averages the repeatedly observed measures for $X_i(t)$ across the $J$ replicates to obtain $\overline{W}_{i\cdot}(t)$. The Naive estimator is based on the first replicate $\{W_{i1}(t)\}$ of the observed measures as a substitute for $X_i(t)$. The Oracle or benchmark estimator uses $X_i(t)$. The Average method provides an ad hoc method for adjusting for biases due to measurement error, while the Naive estimator does not provide any adjustments for the presence of measurement errors. The PACE estimator is obtained using the PACE method proposed by Yao, et al. \citep{yao2005functional}.

We performed three sets of simulations to investigate the performance of the six estimators in response to three different factors, respectively: 1) increasing sample size, with $n \in (100,200, 500, 1,000, 2,000, 5,000)$, $\sigma_x = 3.0$, where $\sigma_x$ indicates the diagonal elements in the covariance matrix for $X_i(t)$, and $\rho_x = 0.50$ (see Table \ref{Table:samplesize}); 2) increasing magnitude of the measurement error with $n=5,000$, $\sigma_x \in (1.5, 2.0, 3.0)$, and $\rho_x = 0.75$ (see Table \ref{Table:ME_scale}); and 3) increasing magnitude of the serial correlation across time points for $X_i(t)$ with $\rho_x \in (0.05,0.25,0.5,0.75)$, $n=5,000$, and $\sigma_x = 3.0$ (see Table \ref{Table:ME_correlation}). We have also conducted another set of simulations to investigate the performance of the MP\_MEM estimator when different numbers of time points are involved in the approximation of the true measurement $X(t)$ when $D \in (3, 4, 5, 6)$ (see Table \ref{Table:ME_MPMEM}). Additionally, we obtained the empirical variance of the functional covariate in the scalar-on-function logistic regression for each method through our simulations and compared them to the variance of the corresponding estimated functional regression coefficient [$\hat{\beta}(t)$] (see Table \ref{Table:Variance}). All simulations were conducted with 1,000 iterations.

We compared the estimators with three measures. Let $\hat{\beta}^{r}(t)$ be the estimator of $\beta(t)$ at the $r$th $(r= 1,\ldots,R)$ simulation replicate and $\overline{\beta}(t)=\frac{1}{R}\sum_{r=1}^{R}\hat{\beta}^{r}(t)$, the averaged estimate over a total of $R=1,000$ replications. Let $\left\{t_{l}\right\}_{l=1}^{n_{grid}}$ be a sequence of equally spaced grid points on $[0,1]$. We define the average squared bias (ABias$^2$) of $\hat{\beta}(t)$, the average sample variance (AVar), and the average integrated mean square error (AIMSE) as 
\begin{eqnarray*}
\text{ABias}^2\{\hat{\beta}(t)\}&=&\frac{1}{n_{grid}}\sum_{l=1}^{n_{grid}}\{\overline{\beta}(t_{l})
-{{{\beta_{true}(t_{l})}}}\}^{2},\\
\text{AVar}\{\hat{\beta}(t)\}&=& \frac{1}{R}\sum_{r=1}^{R}\frac{1}{n_{grid}}\sum_{l=1}^{n_{grid}}\left\{\hat{\beta}^{r}(t_l)-\overline{{\beta}}(t_{l})\right\}^{2}, \text{and} \\
\text{AIMSE}\{\hat{\beta}(t)\}&=&\text{ABias}^2\{\hat{\beta}(t)\}+\text{AVar}\{\hat{\beta}(t)\}, \text{respectively.}
\end{eqnarray*}
Small values of these measures indicate good performance and large values indicate poor performance. 

\subsection{Simulation results}

% \begin{comments}
 %  \subsubsection{Estimation of the covariance function of the measurement error}

%There are two approaches to estimating the covariance function of the measurement error using the repeated measures available on the observed measures of $X_i(t)$. The first method is by directly estimating the covariance matrix of measurement error from the repeated measures available on the observed measures. The second method is through the use of the residuals obtained from implementing the generalized functional linear regression models to estimate the empirical covariance functions. These two approaches produce equivalent results, which were verified in our simulation studies.

%Using the repeated measures available on the observed measures, $W_{ij}(t)$, the covariance function of the measurement errors can be estimated as: 
%\begin{eqnarray*}
%\widehat{\Sigma}_{uu}(t,t') &=& \frac{\sum_{i=1}^n \sum_{j=1}^J\left[W_{ij}(t')-\overline{W}_i(t')\right]\left[W_{ij}(t)-\overline{W}_i(t)\right]}{n (J-1)J}
%\end{eqnarray*}

%While estimates of the empirical covariance functions for $U_{ij}(t)$ may be obtained using the variance functions of $U_{ij}(t) = {W_{ij}(t)} - E\{W_{ij}(t)|X_i(t)\}$, which are functions of the raw residuals from the generalized mixed effects model. Then the variance of measurement error, $U_{ij}(t)$, can be estimated as:

%\begin{eqnarray*}
%\widehat{\Sigma}_{uu}(t,t') &=& \frac{\sum_{i=1}^n \left[U_{ij}(t')-\overline{U}_i(t')\right]\left[U_{ij}(t)-\overline{U}_i(t)\right]}{n}
%\end{eqnarray*}
% \end{comments}

\subsubsection{Impact of sample size}
Table \ref{Table:samplesize} summarizes the results for varying sample sizes. As the sample size increased, the ABias$^2$, AVar, and AIMSE consistently decreased for all the estimators. The MP\_MEM estimator had ABias$^2$ consistently smaller than that of the UP\_MEM, PACE, Average, and Naive estimators and closest to that of the Oracle estimator. The Oracle estimator had the smallest Avar, followed in order by the MP\_MEM, UP\_MEM, Naive, and Average methods. Similarly, the Average method had the highest values for AIMSE while the AIMSE associated with the Oracle estimator was the smallest.  

We observed that ABias$^2$ of the Average estimator is larger than that of the Naive estimator, which is different from what we usually observe in the classical additive measurement error model. However, the measurement error model proposed in this manuscript is different from the classical additive measurement errors model as we allow the observed measurement to have a nonlinear relationship and also relax the assumption on the distribution of measurement error. 
%For the classical additive measurement error model with normally distributed measurement error, maximum likelihood estimation can be obtained in linear regression while it does not always work for more complex models such as logistic regression \citep{carroll:2006}. This explains why we did not observe the usual results that the Average estimator has a smaller bias than the Naive estimator. 

\begin{table}[h] %% Table 1
\caption{The effect of varying sample size on the performance of $\hat{\beta}(t)$ under the six estimators with $n \in (100,200, 500, 1,000, 2,000, 5,000)$, $\sigma_x = 2.0$, and $\rho_x = 0.50$. 
\label{Table:samplesize}}
\centering
\scalebox{0.60}{
\hskip-2.2cm 
\begin{tabular}[t]{r|rrrrrr|rrrrrr|rrrrrr}
\hline
\multicolumn{1}{c} {} &\multicolumn{6}{c} {ABias$^2$} &\multicolumn{6}{c} {AVar} &\multicolumn{6}{c} {AIMSE} \\
\hline
n &  Oracle & MP\_MEM & UP\_MEM &PACE & Average& Naive   &  Oracle & MP\_MEM & UP\_MEM&PACE &Average & Naive  &  Oracle & MP\_MEM & UP\_MEM &PACE & Average& Naive    \\
\hline
100 & 0.0067 & 0.0186 & 0.0290 & 0.2707 & 0.2983 & 0.2650 & 1.8983 & 2.2526 & 2.4473 & 5.4870 & 5.3793 & 5.2546 & 1.9050 & 2.2712 & 2.4763 & 5.7577 & 5.5196 & 5.6776\\

200 & 0.0038 & 0.0113 & 0.0199 & 0.2377 & 0.2581 & 0.2300 & 0.8359 & 0.9814 & 1.0699 & 2.4371 & 2.3411 & 2.2857 & 0.8397 & 0.9927 & 1.0898 & 2.6748 & 2.5158 & 2.5992\\

500 & 0.0006 & 0.0042 & 0.0091 & 0.1790 & 0.1950 & 0.1667 & 0.3613 & 0.4263 & 0.4597 & 1.0800 & 1.0046 & 0.9663 & 0.3618 & 0.4305 & 0.4688 & 1.2590 & 1.1330 & 1.1996\\

1000 & 0.0004 & 0.0030 & 0.0074 & 0.1647 & 0.1776 & 0.1516 & 0.2020 & 0.2381 & 0.2560 & 0.6254 & 0.5512 & 0.5293 & 0.2023 & 0.2412 & 0.2634 & 0.7900 & 0.6808 & 0.7289\\

2000 & 0.0001 & 0.0027 & 0.0071 & 0.1682 & 0.1810 & 0.1560 & 0.1218 & 0.1450 & 0.1550 & 0.3897 & 0.3347 & 0.3226 & 0.1219 & 0.1477 & 0.1621 & 0.5579 & 0.4786 & 0.5157\\

5000 & 0.0001 & 0.0021 & 0.0060 & 0.1573 & 0.1695 & 0.1447 & 0.0628 & 0.0751 & 0.0797 & 0.2080 & 0.1710 & 0.1629 & 0.0629 & 0.0772 & 0.0857 & 0.3653 & 0.3077 & 0.3404\\
\hline

\end{tabular}
}
\end{table}

\subsubsection{Impact of the magnitude of measurement error}
Table \ref{Table:ME_scale} provides the results on varying levels of the magnitude of measurement error. The Oracle estimator consistently had the smallest Abias$^{2}$ and AVar of the six estimators. The MP\_MEM estimator had the next smallest Abias$^{2}$ and AVar values, followed by, in ascending order, the UP\_MEM, PACE, Naive, and Average estimators. The Oracle estimator had the smallest values of Avar and AIMSE while the MP\_MEM estimator had the second smallest values. 

\begin{table}[h] %% Table 2

\caption{The effect of the varying magnitude of measurement error on the performance of $\hat{\beta}(t)$ for the six estimators with $n=5,000$, $\sigma_x \in (1.5, 2.0, 3.0)$, and $\rho_x = 0.75$   
\label{Table:ME_scale}}
\centering
\scalebox{0.60}{
\hskip-2.2cm 
\begin{tabular}[t]{r|rrrrrr|rrrrrr|rrrrrr}
\hline
\multicolumn{1}{c} {} &\multicolumn{6}{c} {ABias$^2$} &\multicolumn{6}{c} {AVar} &\multicolumn{6}{c} {AIMSE} \\
		\hline
			$\sigma_{x}$ &  Oracle & MP\_MEM & UP\_MEM &PACE & Average& Naive   &  Oracle & MP\_MEM & UP\_MEM&PACE &Average & Naive  &  Oracle & MP\_MEM & UP\_MEM &PACE & Average& Naive    \\

		\hline
1.5 & 0.0001 & 0.0015 & 0.0073 & 0.1610 & 0.1757 & 0.1498 & 0.1111 & 0.1282 & 0.1362 & 0.3591 & 0.2815 & 0.2576 & 0.1112 & 0.1297 & 0.1435 & 0.5201 & 0.4075 & 0.4572\\
2.0 & 0.0000 & 0.0024 & 0.0076 & 0.1684 & 0.1796 & 0.1661 & 0.0627 & 0.0740 & 0.0774 & 0.2049 & 0.1618 & 0.1532 & 0.0627 & 0.0764 & 0.0851 & 0.3733 & 0.3193 & 0.3414\\
3.0 & 0.0000 & 0.0047 & 0.0102 & 0.1741 & 0.1820 & 0.1767 & 0.0287 & 0.0355 & 0.0371 & 0.0968 & 0.0747 & 0.0731 & 0.0287 & 0.0402 & 0.0473 & 0.2710 & 0.2498 & 0.2567\\

\hline
\end{tabular}
}
\end{table}

\subsubsection{Impact of the strength of serial correlation across time points}
 Table \ref{Table:ME_correlation} summarizes the impacts of varying the strength of the serial correlation across time points for the functional covariate, $X_i(t)$. The simulation results demonstrate that the biases associated with the Oracle, MP\_MEM, and UP\_MEM methods were all smaller than the bias associated with the PACE method as well as those obtained under the Average and Naive approaches. The strength of correlation of the functional covariate did not affect the estimators obtained under the Oracle and MP\_MEM methods as their ABias$^2$'s remained more stable for increasing values of $\rho_x$. The bias associated with the PACE method is particularly smaller when $\rho_x$ is 0.05 than when the $\rho_x$ values were all larger than 0.05. This result is reasonable because one of the assumptions of the PACE method is discrete (i.e., $\rho_x=0$) measurement error. As the correlation coefficient, $\rho_x$, increased, the Avar and AIMSE of all six estimations decreased. The Oracle estimator had the smallest values of Avar and AIMSE and was followed in decreasing order by the MP\_MEM, UP\_MEM, PACE, Average, and Naive estimators.

\begin{table}[h] %% Table 2

\caption{The effect of increasing magnitude of the serial correlation across time points for $X_i(t)$ on the performance of $\hat{\beta}(t)$ for the six estimators with $\rho_x \in (0.05,0.25,0.5,0.75)$, $n=5,000$, and $\sigma_x = 2.0$.   \label{Table:ME_correlation}}
\centering
\scalebox{0.60}{
\hskip-2.2cm 
\begin{tabular}[t]{r|rrrrrr|rrrrrr|rrrrrr}
\hline
\multicolumn{1}{c} {} &\multicolumn{6}{c} {ABias$^2$} &\multicolumn{6}{c} {AVar} &\multicolumn{6}{c} {AIMSE} \\
		\hline
			$\rho_{x}$ &  Oracle & MP\_MEM & UP\_MEM &PACE & Average& Naive   &  Oracle & MP\_MEM & UP\_MEM&PACE &Average & Naive  &  Oracle & MP\_MEM & UP\_MEM &PACE & Average& Naive    \\

		\hline
0.05 & 0.0001 & 0.0020 & 0.0024 & 0.0857 & 0.1218 & 0.0787 & 0.1111 & 0.1364 & 0.1378 & 2.2091 & 0.2955 & 0.2743 & 0.1111 & 0.1383 & 0.1401 & 2.2948 & 0.3530 & 0.4172\\
0.25 & 0.0001 & 0.0018 & 0.0039 & 0.1323 & 0.1474 & 0.1107 & 0.0826 & 0.0998 & 0.1042 & 0.2554 & 0.2238 & 0.2127 & 0.0827 & 0.1017 & 0.1081 & 0.3877 & 0.3235 & 0.3712\\
0.50 & 0.0001 & 0.0021 & 0.0060 & 0.1573 & 0.1695 & 0.1447 & 0.0628 & 0.0751 & 0.0797 & 0.2080 & 0.1710 & 0.1629 & 0.0629 & 0.0772 & 0.0857 & 0.3653 & 0.3077 & 0.3404\\
0.75 & 0.0000 & 0.0024 & 0.0076 & 0.1684 & 0.1796 & 0.1661 & 0.0627 & 0.0740 & 0.0774 & 0.2049 & 0.1618 & 0.1532 & 0.0627 & 0.0764 & 0.0851 & 0.3733 & 0.3193 & 0.3414\\

\hline
\end{tabular}
}
\end{table}

\subsubsection{Impact of the number of time points involved in MP\_MEM}

Table \ref{Table:ME_MPMEM} provides the results of investigating how the varying number of time points (D) involved to obtain the approximates of the true measurement, $\hat{X}(t)$, impacted the performance of $\hat{\beta}(t)$ for the MP\_MEM approach, while comparing it the other five estimators. Since the changing the value of D did not affect any other approaches except for the MP\_MEM approach, the simulation results for all the other approaches stayed the same as the D increased, while the Abias$^{2}$ and AVar of the MP\_MEM estimator decreased, which suggested that the performance of $\hat{\beta}(t)$ for the MP\_MEM approach gets better when data at more time points are used to approximate the true measurement in MP\_MEM approach. However, the changes between different values of D were not very significant, and the computation time and burden significantly increased when D increased, especially for a larger sample size. For instance, in our simulation study with a sample size of 500, the running time for the MP\_MEM approach when $D = (3, 4, 5, 6)$ are: 2.08, 7.95, 22.06, 50.31 minutes, respectively, using R version 4.3.2 (2023-10-31) on a MacBook Pro (2021) of 32GB memory. In other words, there is a trade-off between the performance of the MP\_MEM approach and the computation burden. One may choose the optimal value of $D$ based on their needs and sample data of interest.

\begin{table}[h] %% Table 2

\caption{The effect of increasing number of time points, $D$, involved in MP\_MEM on the performance of $\hat{\beta}(t)$ for the MP\_MEM estimators with $D \in (3, 4, 5, 6)$, $\rho_x =0.5$, $n=500$, and $\sigma_x = 2.0$.   \label{Table:ME_MPMEM}}
\centering
\scalebox{0.60}{
\hskip-2cm 
\begin{tabular}[t]{r|rrrrrr|rrrrrr|rrrrrr}
\hline
\multicolumn{1}{c} {} &\multicolumn{6}{c} {ABias$^2$} &\multicolumn{6}{c} {AVar} &\multicolumn{6}{c} {AIMSE} \\
		\hline
			$D$ &  Oracle & MP\_MEM & UP\_MEM &PACE & Average& Naive   &  Oracle & MP\_MEM & UP\_MEM&PACE &Average & Naive  &  Oracle & MP\_MEM & UP\_MEM &PACE & Average& Naive    \\

		\hline
3 & 0.0009 & 0.0044 & 0.0092 & 0.1730 & 0.1909 & 0.1650 & 0.3459 & 0.4112 & 0.4440 & 1.0222 & 0.9595 & 0.9270 & 0.3469 & 0.4155 & 0.4532 & 1.1951 & 1.0920 & 1.1504\\

4 & 0.0009 & 0.0042 & 0.0092 & 0.1730 & 0.1909 & 0.1650 & 0.3459 & 0.4102 & 0.4440 & 1.0222 & 0.9595 & 0.9270 & 0.3469 & 0.4144 & 0.4532 & 1.1951 & 1.0920 & 1.1504\\

5 & 0.0009 & 0.0041 & 0.0092 & 0.1730 & 0.1909 & 0.1650 & 0.3459 & 0.4094 & 0.4440 & 1.0222 & 0.9595 & 0.9270 & 0.3469 & 0.4135 & 0.4532 & 1.1951 & 1.0920 & 1.1504\\

6 & 0.0009 & 0.0040 & 0.0092 & 0.1730 & 0.1909 & 0.1650 & 0.3459 & 0.4092 & 0.4440 & 1.0222 & 0.9595 & 0.9270 & 0.3469 & 0.4132 & 0.4532 & 1.1951 & 1.0920 & 1.1504\\
\hline

\end{tabular}
}
\end{table}
\subsubsection{Variance of the estimated functional regression coefficient}
%%%%%%%%%%%%%%%%%%% edition by YL %%%%%%%%%%%%%%%%%%%
Our proposed measurement error model is different from the classical additive measurement errors model because we allow the observed measurement to have a nonlinear relationship, which is determined by the link function $h(\cdot)$, with the true measurement. In our simulation studies, we let $h(\cdot)$ be natural $log(\cdot)$. In classical approaches to adjusting for measurement error in regression models, a bias-variance trade-off is often observed. However, in the simulation results presented, we did not observe the usual bias-variance due to the natural log transformation of the observed measures during the estimation. For a scalar-on-scalar generalized linear regression, we can show the asymptotic relationship between the variance of the covariates and the variance of the estimated parameters. For instance, for a scalar-on-scalar logistic regression, the variance of the estimated coefficients, $\hat{\beta}$, is given by the inverse of the Fisher information matrix, $I(\beta)$. That is $\sigma_{\hat{\beta}}^2=[I(\beta)]^{-1}$, where $I(\beta)=\boldsymbol{X}^T \boldsymbol{\Omega X}$. $\boldsymbol{X}$ is the matrix of covariates and $\boldsymbol{\Omega}$ is a diagonal matrix with the element $\omega_i=p_i\left(1-p_i\right)$, and $p_i$ is the predicted probability from the logistic model. Asymptotically, $\sigma_{\hat{\beta}}^2$ is inversely related to the sample size and the variance of the covariates $\boldsymbol{X}$. Therefore, under the same sample size, a covariate with a larger variance is expected to produce the regression coefficient with a smaller variance. Based on the Delta method, the variance of the Naive covariate used to obtain the Naive estimator in a scalar-on-scalar regression is given by: $\operatorname{Var}[\log (W_{1} +1)] \approx \left(\frac{1}{\overline{W_{\cdot1}}+1}\right)^2 \operatorname{Var}(W_{1})$. Since $\overline{W_{\cdot1}}+1 >> 1$, we can have that $\operatorname{Var}[\log (W_{1} +1)] << \operatorname{Var}(W_1)$ and thus $\operatorname{Var}[\beta_{\log (W_1+1)}] >> \operatorname{Var}[\beta_{W_1}]$. Therefore, after the natural log transformation, the variance covariate used to obtain the Naive regression coefficient estimator is significantly reduced, which will lead to estimated regression coefficients with increased variance. Intuitively, we would expect the same for the scalar-on-function logistic regression. However, demonstrating this analogous asymptotic relationship for the scalar-on-function logistic regression model is more complexn and beyond the scope of this manuscript. Instead, we obtained the empirical variance of the functional covariate in the scalar-on-function logistic regression for each method through our simulations and demonstrated the relationship between the variance of functional covariate and the variance of the corresponding estimated functional regression coefficient [$\hat{\beta}(t)$] numerically. From Table \ref{Table:Variance}, we can see that the variance (Avar) of $\hat{\beta}(t)$ tended to be inversely associated with the variance (Avar) of the corresponding functional covariate, which explains why we did not observe the usual bias-variance trade-off in the previous simulation results since the variance of MP\_MEM and UP\_MEM covariates are higher than that of the PACE, Naive and Average covariates. In summary, we did not observe the usual bias-variance trade-off in our simulation study due to the natural log transformation of the observed measurement.

\begin{table}[h] %% Table 1
\caption{Relationship between the variance of functional covariate and the variance of estimated functional regression coefficient [$\hat{\beta}(t)$] under the six estimators with $n \in (100,200, 500, 1,000, 2,000, 5,000)$, $\sigma_x = 3.0$, and $\rho_x = 0.50$. 
\label{Table:Variance}
}
\centering
\scalebox{0.8}{
\centering
\hskip-1.6cm 
\begin{tabular}[t]{r|rrrrrr|rrrrrr}
\hline
\multicolumn{1}{c} {} &\multicolumn{6}{c} {AVar of Functional Covariate} &\multicolumn{6}{c} {AVar of $\hat{\beta}(t)$} \\
\hline
n &  Oracle & MP\_MEM & UP\_MEM& PACE & Average& Naive &  Oracle & MP\_MEM & UP\_MEM& PACE & Average& Naive\\
\hline
100 & 4.000174 & 3.264172 & 3.304406 & 1.433073 & 1.584075 & 1.698373 & 1.8983 & 2.2526 & 2.4473 & 5.4870 & 5.3793 & 5.2546\\

200 & 4.004995 & 3.265240 & 3.304081 & 1.400254 & 1.584244 & 1.698544 & 0.8359 & 0.9814 & 1.0699 & 2.4371 & 2.3411 & 2.2857\\

500 & 4.000968 & 3.259412 & 3.296737 & 1.381040 & 1.584923 & 1.699299 & 0.3613 & 0.4263 & 0.4597 & 1.0800 & 1.0046 & 0.9663\\
1000 & 4.002925 & 3.260024 & 3.298016 & 1.375071 & 1.585099 & 1.699378 & 0.2020 & 0.2381 & 0.2560 & 0.6254 & 0.5512 & 0.5293\\

2000 & 4.000827 & 3.257178 & 3.294464 & 1.367771 & 1.583912 & 1.698332 & 0.1218 & 0.1450 & 0.1550 & 0.3897 & 0.3347 & 0.3226\\

5000 & 3.999401 & 3.254883 & 3.293308 & 1.363611 & 1.583946 & 1.698281 & 0.0628 & 0.0751 & 0.0797 & 0.2080 & 0.1710 & 0.1629\\
\hline
\end{tabular}}
\end{table}

\section{APPLICATION } \label{apply}

\subsection{Data and Model}

We applied our methods to data 
 from the 2005\textendash2006 cycle of the NHANES to assess the association between wearable device-based physical activity and T2D status while adjusting for age, sex, and race/ethnicity in community-dwelling adults in the U.S. The participants wore ActiGraph uniaxial accelerometers (model \#AM7164) on their hips for at least four days during waking hours except during water-related activities, as the ActiGraph devices were not water-proof. The ActiGraph uniaxial accelerometers (model \#AM7164) registered physical activity intensity and step counts as zero when participants were not wearing them or were sedentary. In this application, we defined the true latent covariate, $X_i(t)$, as the true unknown usual activity counts for a given participant $i$ during wear time $t$. The observed measure for $X_i(t)$ was the step count recorded by the device. Because NHANES did not provide any information to distinguish between non-wear and non-moving time, we defined non-wear time to be when the device-based physical activity measures were recorded as zero for at least 60 consecutive minutes and set the physical activity measures (step counts) to be non-applicable values during the non-wear time. We included data on participants who were older than 20 and had physical activity data for at least four days. We applied sample weights to account for the oversampling of subjects belonging to the Hispanic and black racial groups, per NHANES guidelines \citep{johnson2013national}. The final analytic sample included 2,001 participants, whose Average age was $44 \pm 14$ years. $51\%$ of the participants were female and approximately $14\%$ were black, $13\%$ were Hispanic, $68\%$ were white, and $4\%$ were of another race or ethnicity after adjusting for oversampling from some racial groups. We classified participants who reported being told by a doctor that they have T2D as having a prior diagnosis of T2D, and classified others as Non-T2D participants.

\begin{figure}%[h]
	\centering
	\includegraphics[width=16cm]{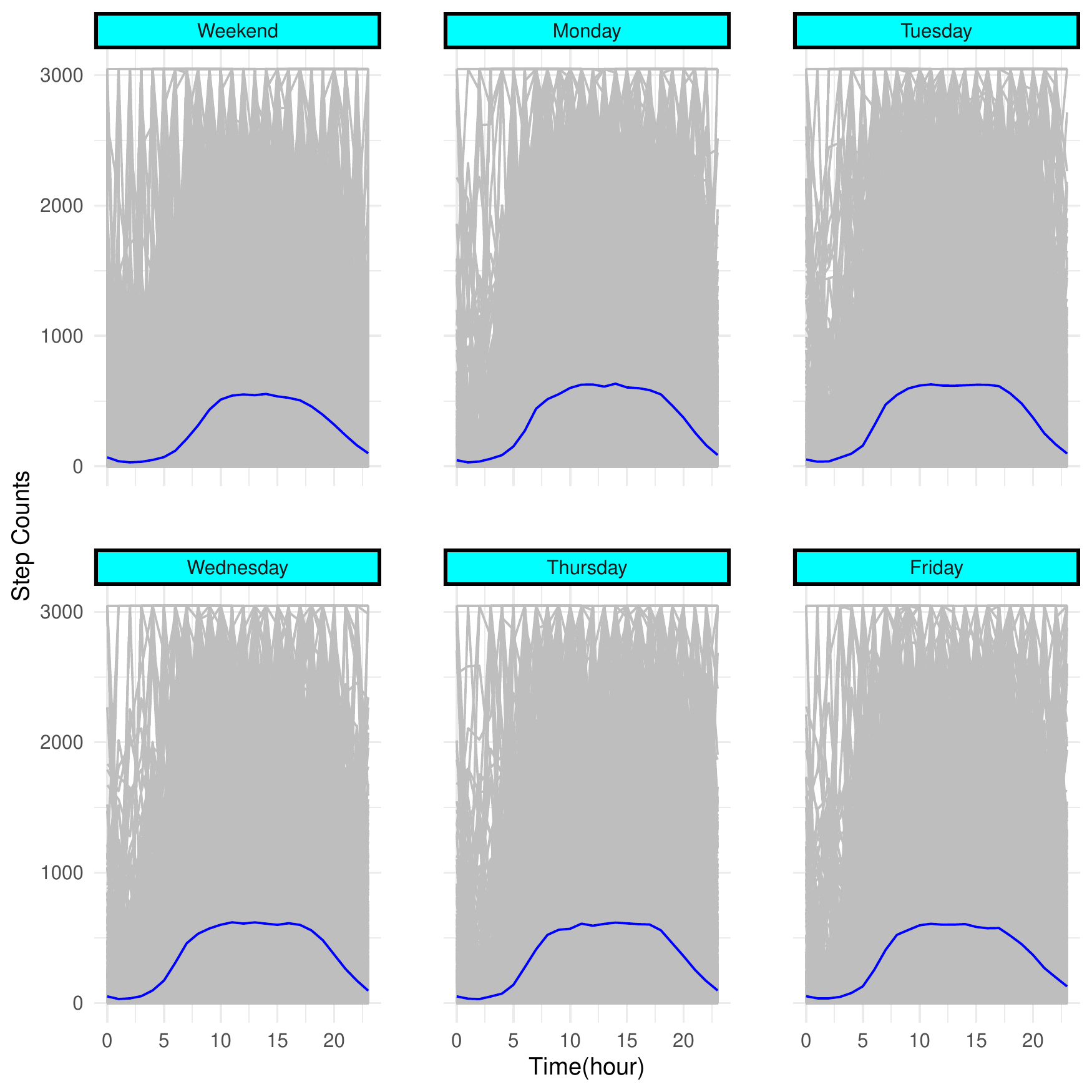}
	\caption{Wearable-device-based step counts from the 2005-2006 NHANES plotted against time (hour) for weekdays and weekends. The grey lines on the background are the trend of step counts over time for each subject and the blue line is the Average step count across all participants over time for each day \label{step2005}}
\end{figure}

\begin{figure}%[h]
	\centering
	\includegraphics[width=16cm]{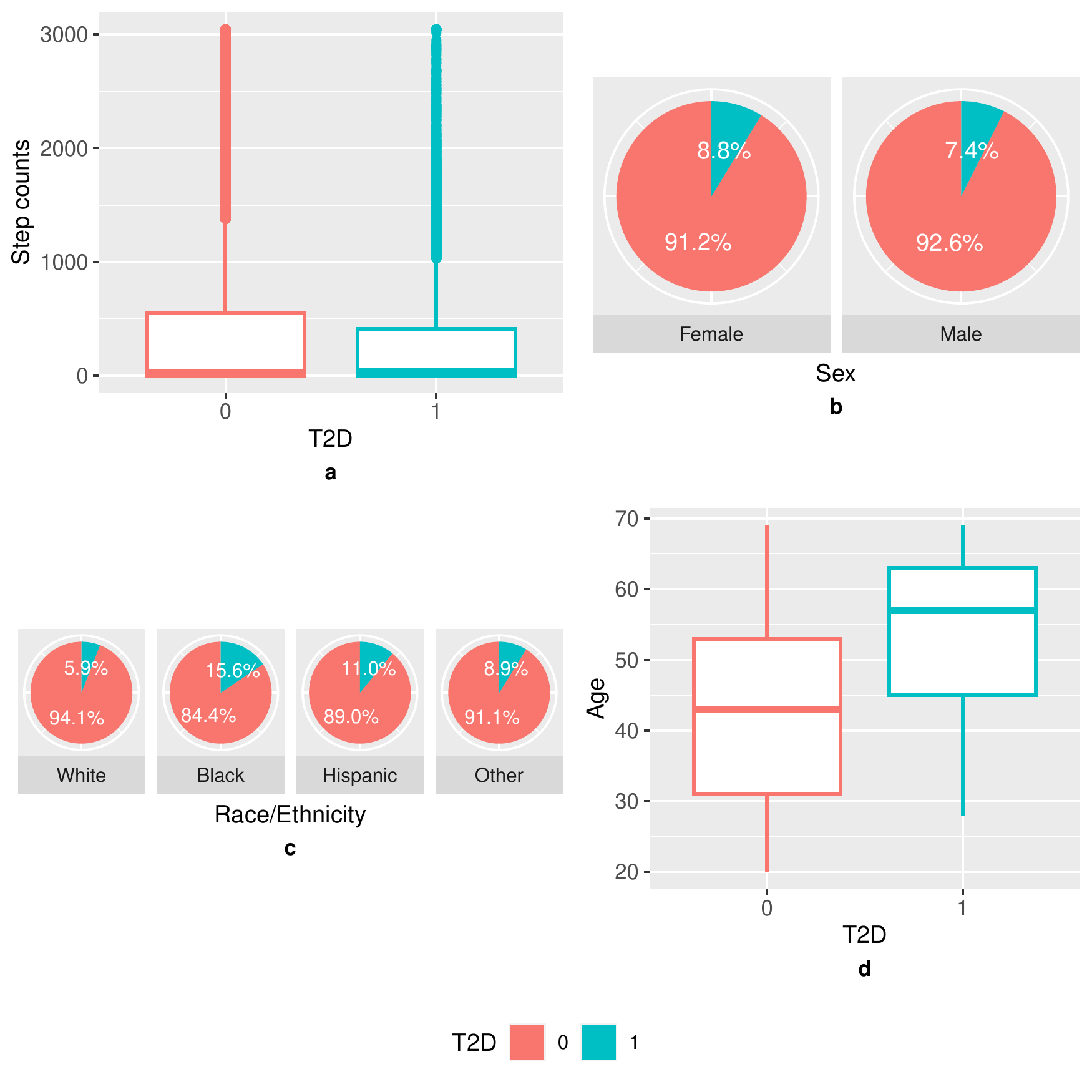}
	\caption{Patterns of T2D status by demographics: (a) boxplot of step counts by T2D status, (b) percentages of females and males with T2D, (c) percentages of participants in each racial group with T2D, and (d) boxplot of age for participants by T2D status. T2D=0: no T2D diagnosis; T2D=1: T2D diagnosis. \label{T2D}}
\end{figure}

 Figure \ref{step2005} illustrates patterns of wearable-device-based step counts for weekdays and weekends by time (hours) over 24 hours. The patterns of step counts on weekdays and weekends were similar. Therefore, we analyzed the data for weekdays and weekends together as a whole. Figure \ref{T2D} illustrates the prevalence of T2D patients by gender and race/ethnicity (Figure \ref{T2D} b, c) as well as the distribution of step counts and age among T2D and non-T2D participants (Figure \ref{T2D} a, d). Participants with T2D tended to have fewer device-based step counts when compared to participants without T2D. The percentage of T2D patients among female participants was slightly higher than that of the male participants. The percentages of blacks, Hispanics, whites, and others who had T2D were $15.6\%$, $11.0\%$, $5.9\%$, $8.9\%$, respectively.

%Figure \ref{T2D} shows a summary of participants in terms of step count, sex, race/ethnicity, and age by T2D status. Participants with T2D tended to have fewer device-based step counts when compared to participants without T2D. Also, similar proportions of females and males had T2D. The percentages of blacks, Hispanics, whites, and others who had T2D were $15.6\%$, $11.0\%$, $5.9\%$, $8.9\%$, respectively. Participants with T2D tended to be older than participants without T2D. 

We considered the following functional logistic regression model with measurement error:
$$
\begin{aligned}
\log \left(\frac{P_i}{1-P_i}\right)&=\int_{0}^{1} \beta(t)X_{i}(t)  dt +  \bZ_i\trans\balpha , \\
\log[E\{W_{ij}(t)|X_i(t)\}] &= X_i(t),\\
W_{ij}(t)|X_i(t) &\sim \text{PP}\{X_i(t)\}
\end{aligned}
$$
where $P_i$ is the probability of being diagnosed with T2D for the $i${th} subject, $\beta(t)$ is the unknown functional regression coefficient, and $\balpha$ is the unknown scalar regression coefficients of the error-free covariates. The $X_{i}(t)$ are the unobservable true usual activity counts, with observed measure, $W_{ij}(t)$, representing device-based step counts prone to measurement error, and $\bZ_i$ are the error-free covariates, including age, sex, and race/ethnicity, for the $i${th} subject. 

\subsection{Application Results}

Table \ref{Table:errorfree} shows the estimated exponentiated regression coefficients of the error-free covariates and their corresponding $95\%$ confidence intervals (CIs) for the MP\_MEM, UP\_MEM, PACE, Average, and Naive estimators of $\beta(t)$. We obtained the $95\%$ CIs via a nonparametric bootstrap with $1,000$ replicates. Correcting for measurement error influenced estimates for some error-free covariates but not others. For example, the statistically significant estimates of the age coefficient were similar for the five estimators. There was no statistically significant difference in the odds of T2D between the male and female participants based on all five estimates. Additionally, the UP\_MEM-, MP\_MEM-, PACE- and Naive-based estimators were smaller than those obtained from the Average-based method. There was an increased odds of T2D among blacks when compared to whites. The estimated odds obtained under the MP\_MEM and UP\_MEM methods were smaller than those obtained under the PACE and Naive approaches. However, the largest estimated difference in odds of T2D between blacks and whites was based on the Naive approach. We also observed an increased odds of T2D among Hispanics when compared with whites, which were not statistically significant. The estimated difference in odds between whites and Hispanics obtained under the MP\_MEM approach was the smallest and it was the largest under the Naive approach, while it was found to be equivalent under the UP\_MEM, PACE, and Average approaches. Finally, we did not observe any difference in the odds of T2D between participants belonging to the other racial/ethnic groups when compared to whites under all the other four methods of estimation, except for the Naive method, which produced a much smaller estimation. The MP\_MEM and UP\_MEM approaches were estimated to be equivalent while the Average-based method was the highest.

\begin{figure}%[h]
	\centering
	\includegraphics[width=16cm]{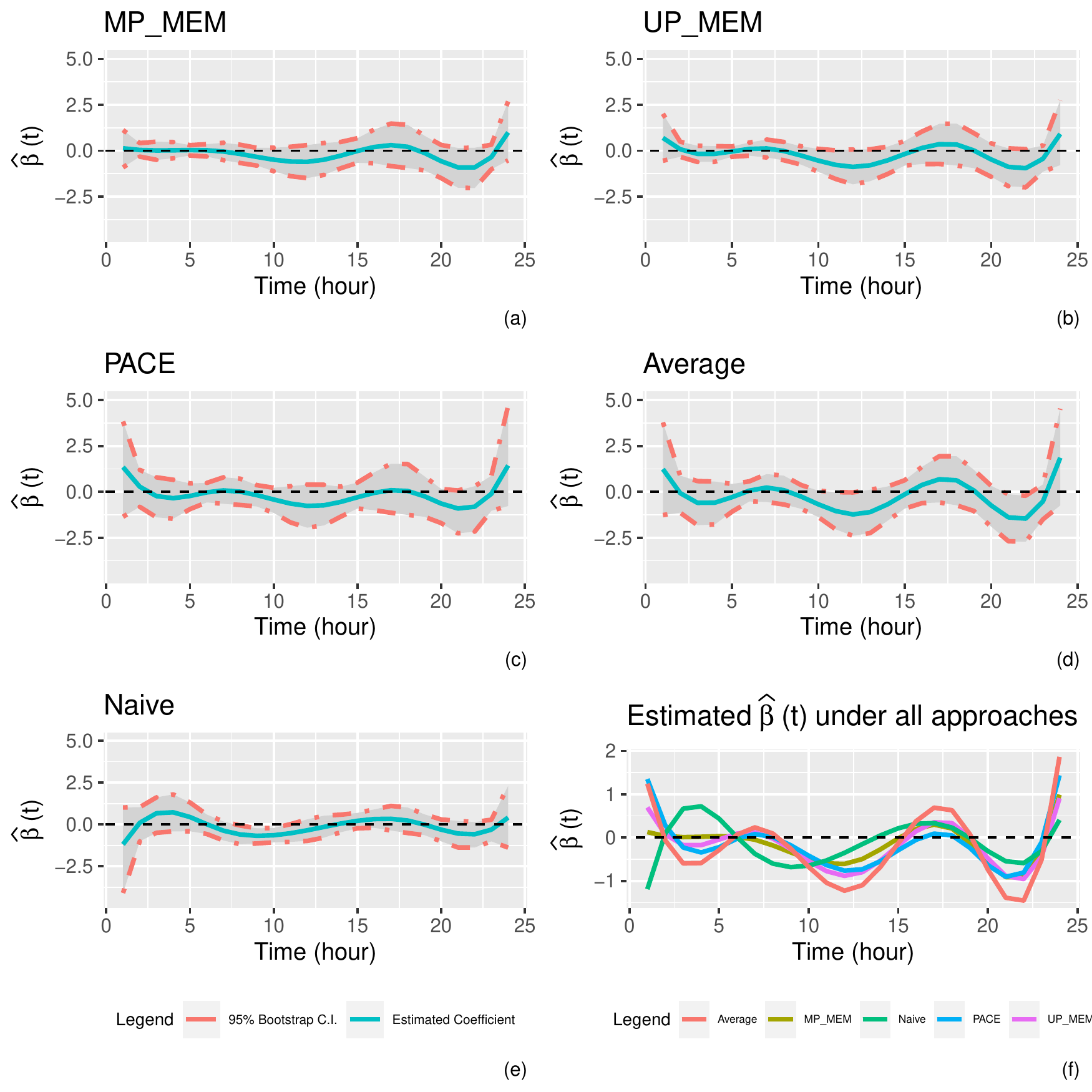}
	\caption{Plots (a-b) of nonparametric $95\%$ point-wise bootstrap confidence intervals (CI) of [$\hat{\beta}(t)$] with each estimator. The shaded areas are the nonparametric $95\%$ point-wise bootstrap confidence intervals (CI) of [$\hat{\beta}(t)$]. The dashed cyan lines represent the $95\%$ point-wise lower and upper bounds of these CIs.
 Plot (e) of comparisons between MP\_MEM, UP\_MEM, Average, and Naive estimates of the coefficients of the association between step count and T2D [$\hat{\beta}(t)$] after adjusting for age, sex, and race in data from the 2005-2006 NHANES.  \label{app_2005}}
\end{figure}

We graphically illustrated the results for the functional coefficient in Figure \ref{app_2005}. The plots demonstrate the $95\%$ non-parametric bootstrap CI as well as the point estimation of the functional coefficient for the association between step count and T2D, [$\hat{\beta}(t)$] under each approach separately [Figure 3(a)-(e)] and the comparisons between the MP\_MEM, UP\_MEM, PACE, Average, and Naive estimators of $\hat{\beta}(t)$ [Figure 3(f)], after adjusting for age, sex, and race. The MP\_MEM, UP\_MEM, and PACE estimators were relatively close to each other and differed from the Average and Naive estimators. The Naive estimator tended to have different time trends when compared to the MP\_MEM, UP\_MEM, PACE, and Average estimators. The MP\_MEM, UP\_MEM, PACE estimators were more attenuated towards the null compared with the Average estimator. In addition, the MP\_MEM estimator was null (i.e., with estimation being 0) between the $0${th} and $6${th} hour, which seems to be more reasonable compared to non-zero estimations since we usually expect to have no physical activities during that time. We observed statistically significant decreases in the odds of T2D and physical activity when physical activity was performed between the $9${th} and $10${th} based on the Naive method. However, this statistical significance was not observed under the MP\_MEM and UP\_MEM methods. Furthermore, the plots indicate that the impact of measurement error in device-based step counts is time-varying.

\begin{table}[h]
 \centering
\caption{Application of five estimators in estimating the associations of the error-free covariates with T2D after correcting for measurement error in device-based step counts in data from the 2005-2006 NHANES. \label{Table:errorfree}}
   
   \scalebox{0.66}{
   \hskip-2.2cm 
	\begin{tabular}{|cc| rr rr rr rr rr |} % [t]
		\hline
		\textbf{Covariates} & & \textbf{MP\_MEM}  & \textbf{$95\%$ C.I.} & \textbf{UP\_MEM}  & \textbf{$95\%$ C.I.} & \textbf{PACE}  & \textbf{$95\%$ C.I.} & \textbf{Average} & \textbf{$95\%$ C.I.} & \textbf{Naive} & \textbf{$95\%$ C.I.}  \\
Intercept && -6.7969 & (-7.8419,-6.0407) & -6.8184 & (-7.8748,-6.0695) & -6.7895 & (-7.8478,-6.0621) & -6.8455 & (-7.8948,-6.0811) & -6.8545 & (-7.8573,-6.077)\\

Age&  & 0.0789 & (0.0651,0.0954) & 0.0792 & (0.0656,0.0958) & 0.0787 & (0.0654,0.0959) & 0.0796 & (0.0661,0.0965) & 0.0802 & (0.0664,0.0967)\\

Sex&&&&&&&&&&& \\
& Female & 1.2305 & (0.7283,1.6993) & 1.2329 & (0.7413,1.6975) & 1.2407 & (0.7563,1.7061) & 1.2472 & (0.7566,1.7138) & 1.2096 & (0.7061,1.6634)\\

Race &&&&&&&&& && \\
& Black & 1.2444 & (0.8386,1.7004) & 1.2396 & (0.8395,1.6977) & 1.2530 & (0.8455,1.6998) & 1.2414 & (0.825,1.6949) & 1.2687 & (0.8569,1.7144)\\

& Hispanic & 0.8185 & (-0.4358,1.6347) & 0.8315 & (-0.4113,1.6542) & 0.8424 & (-0.3587,1.7069) & 0.8356 & (-0.3593,1.6577) & 0.8613 & (-0.3213,1.7293)\\

&Other & 0.0588 & (-0.2797,0.4091) & 0.0559 & (-0.2809,0.4044) & 0.0467 & (-0.2992,0.4075) & 0.0510 & (-0.2946,0.406) & 0.0347 & (-0.3044,0.3711)\\

\hline
\end{tabular}}

\end{table}

\begin{comment}
%%% might add this plots to appendix 
  \begin{figure}%[h]
	\centering
	\includegraphics[width=16cm]{betat_ME_CI_multiple.pdf}
	\caption{Plots of pairwise comparisons between MP\_MEM, UP\_MEM, Average, and Naive estimates of the exponentiated coefficients of the association between step count and T2D [$\exp\{\hat{\beta}(t)\}$] after adjusting for age, sex, and race in data from the 2005-2006 NHANES. The shaded areas are the nonparametric $95\%$ point-wise bootstrap confidence intervals (CI) of [$\exp{\hat{\beta}(t)}$] with the MP\_MEM estimator. The dashed cyan lines represent the $95\%$ point-wise lower and upper bounds of these CIs. \label{app_2005}}
\end{figure}
  
\end{comment}

\section{DISCUSSION}
We developed the multi-level generalized functional linear regression model with functional covariates prone to heteroscedastic errors. These models are suited to massive longitudinal functional data assumed to be error-prone, such as those collected by wearable devices at frequent intervals over multiple days. Additionally, the measurement error component of the model was developed under the multi-level generalized functional linear regression models framework. To date, most approaches to adjusting for biases due to measurement error in functional data analysis are based on the assumption that the observed measures, $W_{ij}(t)$, follow Gaussian distributions. We assumed an exponential family distribution for the observed measures and relaxed the assumption on the probability distribution of measurement error. We implemented functional mixed effects-based methods to adjust for measurement errors biases and allow arbitrary heteroscedastic covariance functions for the measurement errors. To evaluate our proposed methods, we conducted simulations that involved Poisson assumptions for the observed measures and Gaussian error processes for the measurement errors. The functional mixed effects-based methods generally had lower $\text{Abias}^{2}$ than estimators based on the PACE, Average, and Naive methods that do not correct for measurement error explicitly. While the PACE estimator reduces bias more than the Average and Naive estimators, it did not provide any formal adjustment for the serial correlations associated with measurement error in the estimation. The MP\_MEM estimator yielded lower $\text{Abias}^{2}$ and $\text{Avar}$ than the UP\_MEM estimator because MP\_MEM focuses on multiple time points concurrently in the estimation while UP\_MEM is a univariate approach. Overall, the UP\_MEM and MP\_MEM methods performed better than the PACE, Average and Naive methods with increasing sample sizes and varying levels of correlations in the serially observed functional covariate prone to errors. 

The measurement error model we proposed in this manuscript allows the observed measurement prone to measurement error and the true measurement to have a non-linear relationship, which makes the proposed model different from the classical additive measurement error. Such difference leads to different findings than what we usually observe in the classical additive measurement error model. First, we did not observe a bias-variance trade-off in the regression coefficient estimations in our simulations. Second, we observed that the Naive estimator had smaller $\text{Abias}^{2}$ than the Average estimator. 

We illustrated our new methods in analyses of data from the NHANES. We assessed the association between a device-based measure of step count, an observed true usual physical activity counts, and T2D in adults in the U.S. The MP\_MEM and UP\_MEM estimators were very close except for the first 6 hours of the day in our analyses and they demonstrated more adjustment of measurement error compared to the PACE, Average, and Naive estimators. 
%Furthermore, the profiles of the error-adjusted estimators and non-error-adjusted estimators varied across demographic subgroups, suggesting that different subgroups may require separate corrections for measurement error. 

%%%%%%%%%%%%%%%%%%%%%%%%%%%%%%%%%%%%%%%%%%%%%%
%% Support information, if any,             %%
%% should be provided in the                %%
%% Acknowledgements section.                %%
%%%%%%%%%%%%%%%%%%%%%%%%%%%%%%%%%%%%%%%%%%%%%%
\begin{acks}[Acknowledgments]
{\it Conflict of Interest}: None declared.
%The authors would like to thank Dr. Devon Brewer for his constructive feedback and for editing the manuscript. 

\end{acks}

%%%%%%%%%%%%%%%%%%%%%%%%%%%%%%%%%%%%%%%%%%%%%%
%% Funding information, if any,             %%
%% should be provided in the                %%
%% funding section.                         %%
%%%%%%%%%%%%%%%%%%%%%%%%%%%%%%%%%%%%%%%%%%%%%%
\begin{funding}
This research was supported by the National Institute of Diabetes and Digestive and Kidney Diseases Award 1R01DK132385-01. This research was also supported in part by Lilly Endowment, Inc., through its support for the Indiana University Pervasive Technology Institute.

\end{funding}

%%%%%%%%%%%%%%%%%%%%%%%%%%%%%%%%%%%%%%%%%%%%%%
%% Supplementary Material, including data   %%
%% sets and code, should be provided in     %%
%% {supplement} environment with title      %%
%% and short description. It cannot be      %%
%% available exclusively as external link.  %%
%% All Supplementary Material must be       %%
%% available to the reader on Project       %%
%% Euclid with the published article.       %%
%%%%%%%%%%%%%%%%%%%%%%%%%%%%%%%%%%%%%%%%%%%%%%

%%  imsart-nameyear.bst  will be used to                   %%
%%  MR numbers will be added by VTeX.                      %%
%%                                                         %%
%%  Use \cite{...} to cite references in text.             %%
%%                                                         %%
%%%%%%%%%%%%%%%%%%%%%%%%%%%%%%%%%%%%%%%%%%%%%%%%%%%%%%%%%%%%%

%% if your bibliography is in bibtex format, uncomment commands:
%\bibliographystyle{imsart-nameyear} % Style BST file
      % Bibliography file (usually '*.bib')

%%Harvard (name/date)
\bibliographystyle{SageH}
%%Vancouver (numbered)
%\bibliographystyle{SageV}
\bibliography{cqr_fmem_YL.bib} 
\end{document}